\documentclass{ws-ijmpe}
\usepackage[super,compress]{cite}
\usepackage{color}

\newcommand {\bea}{\begin{eqnarray}}
\newcommand {\eea}{\end{eqnarray}}
\newcommand {\be}{\begin{equation}}
\newcommand {\ee}{\end{equation}}

\begin{document}

\def\({\left(}
\def\){\right)}
\def\[{\left[}
\def\]{\right]}

\def\Journal#1#2#3#4{{\it #1} {\bf #2}, (#3) #4}
\def\RPP{{Rep. Prog. Phys}}
\def\PRC{{Phys. Rev. C}}
\def\PRD{{Phys. Rev. D}}
\def\ZPA{{Z. Phys. A}}
\def\NPA{{Nucl. Phys. A}} 
\def\JPG{{J. Phys. G }}
\def\PRL{{Phys. Rev. Lett}}
\def\PRpt{{Phys. Rep.}}
\def\PLB{{Phys. Lett. B}}
\def\AP{{Ann. Phys (N.Y.)}}
\def\EPJA{{Eur. Phys. J. A}}
\def\NP{{Nucl. Phys}}  
\def\RMP{{Rev. Mod. Phys}}
\def\IJMPE{{Int. J. Mod. Phys. E}}
\def\APJ{{Astrophys. J}}  
\def\APJS{{Astrophys. J. Suppl. Ser.}}  
\def\AA{{Astron. Astrophys}}
\def\MPLA{{Mod. Phys. Lett. A}}

\markboth{A. Sulaksono, Naosad Alam and B. K.  Agrawal}
{Core-crust transition properties  of neutron stars}

\catchline{}{}{}{}{}

\title{Core-crust transition properties  of neutron stars within
systematically varied extended relativistic mean-field model}

\author{\footnotesize A. Sulaksono}

\address{Departemen Fisika, FMIPA, Universitas Indonesia,\\
Depok, 16424, Indonesia}

\author{\footnotesize Naosad Alam}

\address{Saha Institute of Nuclear Physics, Kolkata 700
064, India}

\author{\footnotesize B. K. Agrawal}

\address{Saha Institute of Nuclear Physics, Kolkata 700
064, India}

\maketitle

\begin{history}
\received{(received date)}
\revised{(revised date)}
\end{history}

\begin{abstract}
The model dependence  and the symmetry energy dependence of the core-crust
transition properties for the neutron stars are studied using three
different families of systematically varied extended relativistic mean
field model.  Several forces within each of the families are so considered
that they yield wide variations in the values of the nuclear symmetry
energy $a_{\rm sym}$ and its slope parameter $L$ at the saturation
density. The core-crust transition density is calculated using a method
based on  random-phase-approximation.  The core-crust transition density
is strongly correlated, in a model independent manner,  with the symmetry
energy slope parameter evaluated at the saturation density. The pressure
at the transition point dose not show any meaningful correlations with
the symmetry energy parameters at the saturation density.  At best,
pressure at the transition point is correlated with the symmetry
energy parameters and their linear combination  evaluated at the
some sub-saturation density.  Yet, such correlations might not be model
independent. The correlations of core-crust transition properties with
the symmetry energy parameter are also studied by varying the symmetry
energy within a single model.  The pressure at the transition point is
correlated once again with the symmetry energy parameter at the sub-saturation density.   
\end{abstract}
\ccode{PACS numbers:26.60.Gj; 21.30.Fe; 26.60.Kp}
\section{Introduction}
\label{sec_intro}
The matter in the outer core of the neutron stars (NS)
become unstable, against the density fluctuations,
below a particular density  so called core-crust transition density.
The knowledge of the core-crust transition properties in NS matter  is
very important in understanding the pulsar glitches, crust relaxation
in cooling and  accreting neutron stars and asteroseismology from giant
magnetar flares \cite{Lattimer2012}.  The values of core-crust transition
density and the corresponding pressure depend crucially on the behaviour
of nuclear symmetry energy around  the sub-saturation densities.
However, density dependence of the nuclear symmetry energy  is 
known with large uncertainties.  Some progress in this direction has been
made in the last few years \cite{Fur2002,Steiner2005,Cente2009,Warda2009,
Pie1,Agra2012,Roca-Maza13}.  In the mean while, several investigations
are carried out to study the effects of the variations in the symmetry
energy on the core-crust transition properties.  The variations in the
symmetry energy were achieved either within a single model or by using
large set of randomly selected  models of different types.  The core
crust transition density is found to be strongly correlated with the
various symmetry energy parameters evaluated at the saturation density.
But, results for the correlations between the pressure at the transition
point and the symmetry energy obtained from different investigations
are at variance.

One usually  considers the dependence of the core-crust transition
properties on various nuclear symmetry energy parameters, like, symmetry
energy coefficient $a_{\rm sym}$ and the slope parameter $L$; the later
characterizes the density dependence of the symmetry energy.  It has
been established that the core-crust transition density $\rho_t$ is well
correlated with $L$. However, the actual link between the pressure $P_t$,
at the transition density,  and $L$ is rather uncertain.  The calculations
in Ref. \cite{Mousta2010} were performed within a  density dependent
point coupling (DD-PC) model.  Only a single accurately calibrated
DD-PC model was used to study the correlations of the core-crust
transition properties with the various quantities associated with  the
nuclear matter.  The iso-scalar and iso-vector properties associated with
nuclear matter were varied by changing  the model parameters around their
optimal values.  In Ref. \cite{FP2012}, a covariance analysis based on
a single relativistic mean field (RMF) model was employed to study the
correlations of various neutron star properties with the neutron-skin
thickness in $^{208}$Pb nucleus which is  strongly correlated with the
symmetry energy slope parameter $L$ at the saturation density. In Ref
\cite{Xu2009},  several modified Gogny and commonly used Skyrme Hartree
Fock (SHF) models were used to study the correlations between the core
crust transition properties and the nuclear  symmetry energy parameters.
Both the $\rho_t$ and $P_t$ are found  to be strongly correlated with
$L$ in Refs.  ~\cite{Mousta2010,FP2012,Xu2009}.  The similar studies were
performed in Ref. \cite{Ducoin2011}, but, using appropriately calibrated
several RMF and SHF models.  The $\rho_t$ - $L$ correlations were found
to be strong, while, link between   $P_t$ and $L$ was found to be 
sensitive to the models employed.  The lack of $P_t - L$ correlations
are attributed to the delicate balance between the contributions from the
higher order terms and the shift in the transition density with $L$. The
contributions from the higher order terms are model dependent.  The $P_t$
seemed to be better correlated with appropriate linear combination of
$L$ and the symmetry energy curvature parameter $K_{\rm sym}$, both
of these symmetry energy parameters  were evaluated at a sub-saturation
density.  In particular, the $P_t$ are found to be correlated with $L -
0.343 K_{\rm sym} $ with $L$ and $K_{\rm sym}$ are evaluated at $\rho =
0.1\text{ fm}^{-3}$ \cite{Ducoin2011,Newton2013}.  
It is  quite unclear, why some of the investigations yield strong
correlations between $P_t$ and the various symmetry energy parameters,
while, the lack of such correlations are found  in other studies. 
In order to understand better the dependence of the core-crust transition
properties on the model used or on the various symmetry energy parameters,
it is highly desirable to use different models, with their parameters
varied systematically \cite{Klupfel09} to yield wide variations in the
properties of symmetric and asymmetric nuclear  matter.

In the present work, we investigate the  correlations  between
core-crust transition properties  and the density dependence of the
nuclear symmetry energy   for the NS matter using three different
families of  extended relativistic mean-field (ERMF) model.  We consider
several parameterizations for each of the families of the models which
were obtained by systematic variations in such  way that they yield
wide variations in the values of $a_{\rm sym}$ and its slope parameter
$L$ at the saturation density ~\cite{BA2010,DKA2007}.  The ERMF model
includes the contributions from self- and mixed-interaction terms for
isoscalar-scalar $\sigma$,  isoscalar-vector $\omega$ and isovector-vector
$\rho$ mesons up to the quartic order  \cite{BA2010,KS2009}.  The presence
of $\sigma - \rho$ and $\omega - \rho$ mixed interaction terms might alter
the correlation of various core-crust transition properties to the density
dependence of the nuclear symmetry energy.  Such investigations are not
performed previously in detail \cite{FP2012,Mousta2010,Xu2009,Ducoin2011}.
The transition density is calculated using  the relativistic random phase
approximation (RPA) method. For sake of comparison, we also present some
results for core-crust transition density and the corresponding pressure
obtained by commonly used RMF parameter sets such as NL3, FSU, GM1
and TM1.  We also compare our results with those obtained  from dynamical
and thermo-dynamical methods.  

This paper is organized as follows. In Sec~\ref{sec_CCT}, we present the
method to calculate the $\rho_t$. The brief review of ERMF model is given
in Sec.~\ref{sec_rmf}.  In Sec. \ref{sec_SV} we discuss in some detail
the choices for the different families of the systematically varied ERMF
models. The results obtained for the core-crust transition properties
using different families of the ERMF models are presented in Sec.
\ref{sec_NSM}. Finally, main  conclusions are drawn in  Sec.~\ref{conclu}.

\section{CORE-CRUST TRANSITION} \label{sec_CCT}
Determination of the $\rho_t$ of the star is not easy task in general
due to a very complicated structure of the star inner crust. However,
the critical density at which the uniform matter of NS matter becomes
unstable to a small density fluctuation at low densities, can be used
as a good approximation for $\rho_t$ of NS.
There are 3 methods used widely in literature to study the instability due to
small density fluctuations in low density matter. These are,
the thermo-dynamical
method~\cite{Mousta2010,Xu2009,Ducoin2011,Kubis2007,Lattimer2007,Li2002},
the relativistic~\cite{Nilsen1991,Provindencia2006,Avancini2010},
non-relativistic~\cite{Xu2009,Ducoin2011,Pathick1995,Douchin2000,Ducoin2007}
dynamical methods and  the RPA method based on Green function
formalism~\cite{Lim1989,HP2001,Carr,AS2009}. The thermo-dynamical method
requires matter to fulfill not only the mechanical but also the chemical
stability conditions\cite{Kubis2007,Xu2009}:
\bea
-{\(\frac{\partial P}{\partial v}\)}_\mu &>& 0\nonumber\\
-{\(\frac{\partial \mu}{\partial q_c}\)}_v &>& 0,
\eea 
where $v$ and $ q_c$ are the volume and charge per baryon
number, while $P$ is the pressure and $\mu = \mu_n - \mu_p$
is the difference between  chemical potentials for the neutrons and
protons. Otherwise, the matter becomes unstable. For dynamical
models, the instability region of matter can be located by
examining when the convexity of the free-energy curvature matrix is
violated~\cite{Pathick1995,Douchin2000,Ducoin2007,Nilsen1991,Provindencia2006}.
It was shown by Xu {\it et. al}~\cite{Xu2009} that the thermodynamic
stability is the limit of the non-relativistic dynamical model as $k$
$\rightarrow$ 0 (long-wavelength limit) when the Coulomb interaction is
neglected.   The relativistic RPA method, requires longitudinal dielectric
functions $\varepsilon_L$ $>$ 0  when time component of four momentum
$q_0$ =0 to ensure the stability conditions at low-density region. On the
other hand, it was known for some times~\cite{Nilsen1991,Provindencia2006}
that instabilities predicted by the relativistic dynamical method within
Landau-Vlasov formalism is indeed equivalent to those of relativistic
RPA method.

In this work, we choose relativistic RPA method to calculate core-crust
transition density and pressure of NS matter. Based on this method,
the transition from the core to inner crust in the NS matter takes
place at the largest density for which the following condition has a
solution~\cite{Carr}
\be
\varepsilon_L = {\rm{det}} \[1-D_L(q) \Pi_L(q,q_0=0)\] \le 0 . \label{eq:det}
\ee 

In Eq.~(\ref{eq:det}) $q_0$ is the time component of the four-momentum
transfer  $q^\mu=(q_0,\vec{q}\,)$ and $q=|\vec{q}\,|$. The explicit
form of each element in the longitudinal meson propagator and
longitudinal polarization matrices $D_L(q)$ and $\Pi_L$  can be seen in
Refs~\cite{Carr,AM2006,MA2008}. Note, in addition to standard $\gamma$,
$\omega$, $\sigma$ and $\rho$ propagators, the matrix $D_L(q)$ contains
the contributions from mixed propagators  due to the presences of nonlinear
mixed-interaction terms between various mesons  in the ERMF model.
These propagators are determined from the quadratic fluctuations around
the static solutions that are generated by the second derivatives of
the energy density of matter \cite{AM2006,MA2008}.

\section{EXTENDED RELATIVISTIC MEAN-FIELD MODEL}
\label{sec_rmf}
We consider the NS matter which are composed  of nucleons in
$\beta$ equilibrium.  The dense matter in the core of the NS 
matters can be  described by an effective Lagrangian density for the
nucleons interacting through the exchange of $\sigma$, $\omega$ and $\rho$
mesons. The leptons, as required to fulfill the $\beta$-equilibrium
and charge neutrality conditions, are assumed to be non-interacting.
The Lagrangian density for the ERMF model can be written as

 \be
{\mathcal L} = {\mathcal L}^{\rm lin} +{\mathcal L}^{\rm nlin}_{std}+
 {\mathcal L}^{\rm nlin}_{mix} +{\mathcal L}_{L}  . \label{eq:nuclag}
 \ee where the 
 \bea
 {\mathcal L}^{\rm lin} &=&  \sum_{J=n,p} \overline{\Psi}_{J}[i\gamma^{\mu}\partial_{\mu}-(M-g_{\sigma} \sigma)\nonumber\\&-&(g_{\omega }\gamma^{\mu} \omega_{\mu}+\frac{1}{2}g_{\mathbf{\rho}}\gamma^{\mu}\tau .\mathbf{\rho}_{\mu})]\Psi_{J}
 +\frac{1}{2}(\partial_{\mu}\sigma\partial^{\mu}\sigma-m_{\sigma}^2\sigma^2)\nonumber\\
 &-&\frac{1}{4}\omega_{\mu\nu}\omega^{\mu\nu}+\frac{1}{2}m_{\omega}^2\omega_{\mu}\omega^{\mu}
-\frac{1}{4}\mathbf{\rho}_{\mu\nu}\mathbf{\rho}^{\mu\nu}+\frac{1}{2}m_{\rho}^2\mathbf{\rho}_{\mu}\mathbf{\rho}^{\mu},\nonumber\\
\label{eq:eq3}
 \eea
is a function of the kinetic
terms of nucleons; $\sigma$,  $\omega$ and  $\rho$ mesons; and the
corresponding linear interaction terms of nucleons.
The ERMF model is introduced for the first time by Furnstahl {\it et al}
\cite{FST96}. The  Lagrangian densities for  non-linear
self interaction terms which are usually used in standard RMF model can
be written as
 \begin{eqnarray}
{\mathcal L}^{\rm nlin}_{std}&=&-\frac{\kappa_3}{6M} g_\sigma m_\sigma^2
\sigma^3-\frac{\kappa_4}{24M^2} g_\sigma^2 m_\sigma^2 \sigma^4\nonumber\\
&+&\frac{1}{24} \zeta_0 g_\omega^2 {(\omega_\mu \omega^\mu)}^2,
\label{eq:NLnuclagNM} 
 \end{eqnarray}
where $\sigma$,  $\omega$ and  $\rho$
are the mesons field while $g_\sigma$,  $g_\omega$ and  $g_\rho$ are
their corresponding coupling constants. $m_\sigma$,  $m_\omega$,
$m_\rho$ and $M$ are the masses  for $\sigma$, $\omega$, $\rho$
and nucleons respectively. $\kappa_3$, $\kappa_4$, $\zeta_0$ are the
standard non-linear self interaction parameters. The first two terms
in Eq.~(\ref{eq:NLnuclagNM})  were introduced for the first time
by Boguta-Bodmer~\cite{BB77}. Inclusion of these terms provide more
quantitative description of nuclear matter and finite nuclei properties
than those predicted by simple linear RMF model. One of the significant
effects produced by this terms in nuclear matter is softening the
nuclear incompressibility. However, in general, the incompressibility
predicted by standard RMF model depends sensitively also on the choice
of fitting protocol and observable. For example, NL3~\cite{Lala97}
predicts larger but NL-Z2~\cite{Bender}  smaller incompressibility
compared to the experimental value~\cite{Garg,Li07}. From theoretical
point of view, the positive value of $\kappa_4$ is more favorable,
otherwise the energy spectrum has no lower bound and instabilities
in nuclear matter equation of state (EOS) and finite systems may
occur~\cite{AS2009,Estal}. The third term in  Eq.~(\ref{eq:NLnuclagNM})
was introduced by Sugahara-Toki~\cite{Toki} to overcome this problem.
At high density, the quartic vector isoscalar nonlinear parameter
$\zeta_0$ has effect to  bring down the vector potential and makes
the EOS softer~\cite{Estal,Estal2} and this term plays also crucial
role for transversal stability of nuclear matter due to particle hole
excitations~\cite{AS2009}.  However, in standard RMF model,
too large and unnatural value of $\zeta_0$ can not be avoided to
reach acceptable soft SNM EOS. It is reported that by tuning
the $\zeta_0$ one can generate different limiting neutron-star
masses without too much modifying the behavior of the EOS around
$\rho_0$~\cite{Mulser96,Fatt2012}. It is worthwhile to note that
isovector sector of standard RMF model relies only on just a single
coupling constant $g_{\rho}$, which is usually fixed by the binding
energies of asymmetric nuclei.  This  leaves no way to adjust
the isovector properties, like, neutron-skin thickness in $^{208}$Pb nucleus.
Any attempt to adjust the parameters of the standard RMF model  to
accommodate various iso-vector observables may require compromise with
the quality of the fit to the well known bulk properties of the
finite nuclei.  The contributions of $\omega-\rho$ mixed interaction,
as given by fourth term in Eq.  (\ref{eq:NLnuclagMX}), makes the RMF
model more flexible \cite{HP2001}. The ERMF model contains some additional
mixed interaction terms  which are as follows,
 \bea
{\mathcal L}^{\rm nlin}_{mix}&=&\frac{\eta_1}{2M} g_\sigma
m_\omega^2 \sigma (\omega_\mu \omega^\mu)+\frac{\eta_2}{4M^2}
g_\sigma^2 m_\omega^2 \sigma^2 (\omega_\mu \omega^\mu)\nonumber\\
&+&\frac{\eta_\rho}{2M} g_\sigma m_\rho^2 \sigma (\rho_\mu
\rho^\mu)+\frac{\eta_{ 1 \rho}}{4M^2} g_\sigma^2 m_\rho^2 \sigma^2
(\rho_\mu \rho^\mu)\nonumber\\&+&\frac{\eta_{2\rho}}{4M^2}
g_\omega^2 m_\rho^2(\omega_\mu \omega^\mu)(\rho_\nu \rho^\nu),
\label{eq:NLnuclagMX} 
 \eea 
 where $\eta_1$, $\eta_2$ are the nonlinear
isoscalar mixed-interaction parameters while  $\eta_\rho$, $\eta_{ 1
\rho}$ and $\eta_{ 2 \rho}$ are the  nonlinear isovector mixed-interaction
parameters. The presence of nonzero $\eta_1$ and  $\eta_2$ parameters
in ERMF model provides more freedom to adjust the $\zeta_0$ into the
desired value but still retain the positiveness of  $\kappa_4$
~\cite{Estal,Estal2}. The presence of $\eta_{\rho}$,  $\eta_{1
\rho}$ and $\eta_{2 \rho}$ in this model provides more freedom for
controlling behavior of the EOS of asymmetric nuclear matter not only
at high but also at low densities.  In-particular, this   nonlinear
mixed-interaction terms enable one to vary the density dependence of
the symmetry energy coefficient and the neutron skin thickness in finite
nuclei over a wide range without affecting too much the other properties
of finite nuclei~\cite{BA2010,Fur2002,TSil2005}.  The contributions of the
$\rho$ meson self-coupling is not included in the present work. It affects
the properties of the asymmetric nuclear matter only marginally,  because,
the expectation value of the $\rho$ meson field is orders of magnitude
smaller than those of other mesons involved \cite{BA2010,KS2009}.  
The last term of Eq. (\ref{eq:nuclag})  ${\mathcal L}_{L}$ is the Lagrangian
density for non-interacting leptons which can be written as,
 \be
{\mathcal L}_{L}= \sum_{l=e^-, \mu^-, \nu_e, \bar{\nu}_\mu }\overline{\Psi}_l[i\gamma^{\mu}\partial_{\mu}-M_l]\Psi_l.
 \ee

\section{Choice for the systematically varied parameterizations}
\label{sec_SV}
We study the core-crust transition density and the corresponding pressure
in the NS for the three different families of the  ERMF models obtained
in Ref.  \cite{DKA2007}.  These different families correspond to different
choices of the coupling strength, $\zeta_0$, for the self-interaction
of the $\omega$-mesons (Eq. (\ref{eq:NLnuclagNM})).  The value of
$\zeta_0$ were considered to be  $\zeta_0 = 0.0, 0.03g_\omega^2$ and
$0.06g_\omega^2$.  For each of the family, the remaining parameters of
the model were systematically varied to yield different values of the
neutron-skin thickness in $^{208}$Pb nucleus ~\cite{BA2010,DKA2007}.
In other words, for a given $\zeta_0$, the remaining parameter of the
model were optimized using exactly same set of the protocol  except
for the neutron-skin thickness in $^{208}$Pb. The fitting protocol
comprised of  the experimental data for the total binding energies
and  charge rms radii for many closed shell normal and exotic nuclei
\cite{DKA2007}.  The value of neutron-skin thickness in $^{208}$Pb
nucleus was also considered one of the fit data.  The   value of
neutron-skin thickness in $^{208}$Pb was varied over a wide range of
$0.16 - 0.28\text{ fm}$ as it is not yet well constrained.  In total,
there are twenty-one parameter sets, seven parameter sets for each of
the families of the ERMF model corresponding to different values of
neutron-skin thickness in $^{208}$Pb. These parameter sets were named
as BSR1, BSR2,...,BSR21\cite{DKA2007,BA2010}.  The various properties
of the symmetric nuclear matter associated with the BSR1 - BSR21 forces
lie in a narrow range.  For instance, the binding energy per nucleon
for the symmetric nuclear matter $B/A = 16.11\pm0.04$ MeV, the nuclear
matter incompressibility $K=230.24\pm 9.80$ MeV, the nucleon effective
mass $M^*/M=0.605\pm0.004$ and the saturation density  $\rho_{0} = 0.148
\pm0.003$ fm$^{-3}$.  The quality of the fit to the bulk properties of
the finite nuclei are also nearly the same for all the BSR forces; the rms
errors on the total binding energy and the  charge radii are 1.5 - 1.8 MeV
and $0.025 - 0.04$fm, respectively,  for the nuclei considered in the fit.
Hereafter, the parameter sets BSR1 - BSR7 with $\zeta_0 = 0$, BSR8 -
BSR14 with $\zeta_0 = 0.03g_\omega^2$ and BSR15 - BSR21 with $\zeta_0 =
0.06g_\omega^2$, will be referred to as F1, F2 and F3 families of the ERMF
models, respectively.  The maximum mass for the NS for these three families
of interaction lie in the range of 1.7 - 2.4 $M_\odot$. The highest
(lowest) values of maximum mass are obtained for F1(F3) families.  
The variation in the maximum mass of the neutron star across the families
is predominently due to the  change in the values for the self-coupling of
the $\omega$-mesons. The maximum mass increases only by $\sim 0.03M_\odot$
is due to the change  in the density dependence of the symmetry energy
caused by the increase in neutron-skin thickness from 0.16 to 0.28
fm for the $^{208}$Pb nucleus. As the
neutron-skin thickness of $^{208}$Pb nucleus is strongly correlated with
the symmetry energy slope parameter \cite{Cente2009}, the different
families of systematically varied parameterizations can be  used to
assess the model dependence and the symmetry energy dependence on the
core-crust transition properties.  For the sake of comparison, we also
consider  commonly used RMF parameterizations  such as NL3~\cite{Lala97},
FSU~\cite{Pieka2}, TM1~\cite{Toki} and GM1~\cite{GM91}.
To this end,  we would like to mention that the predictions for
the finite nuclei and nuclear matter around saturation density for the
non-linear RMF model considered here are   more or  less the same as
those for the other varient, like, point coupling and  density dependent
meson exchange models \cite{Typel99,Vretenar05,Burvenich02}. We have
considered the RMF model which includes cross-coupling between various
mesons and the self-coupling of $\omega$-mesons in addition to the
conventionally present cubic and quartic terms for the self-coupling
of the $\sigma$-mesons.  The results for such RMF models \cite{Estal}
are consistant with the trends obtained by Dirac-Brueckner-Hartree-Fock
calculations at densities away from the saturation region.

\section{Core-crust transition properties in NS} 
\label{sec_NSM}
The values of the core-crust transition density are calculated
in the  present work using the RPA method as described briefly
in Sec. \ref{sec_CCT}.  
To facilitate the discussions,  let us consider few definitions.
The density dependent symmetry energy $a_{\rm sym}(\rho)$, the
slope parameter $L(\rho)$ and curvature parameter $K_{\rm sym}(\rho)$
are defined as,
\begin{eqnarray} \label{eq:asym}
a_{\rm sym}(\rho)=\frac{1}{2}\left .\frac{d^2E(\rho,\delta)}{d\delta^2}\right
|_{\delta=0},\\
L(\rho)=3\rho \frac{d a_{\rm sym}(\rho)}{d\rho} ,\\
K_{\rm sym}(\rho)=9\rho^2 \frac{d^2 E_{\rm sym}(\rho)}{d\rho^2},
\end{eqnarray}
where, $E(\rho,\delta)$ is the energy per nucleon at a given density
$\rho$ and asymmetry $\delta=(\rho_n - \rho_p)/\rho$.  We use
Eq. (\ref{eq:asym}) to calculate the symmetry energy, while one often
employs the parabolic approximation and evaluates the symmetry energy as,
\begin{equation}
a_{\rm sym} (\rho) \approx E(\rho, \delta = 1) - E (\rho, \delta = 0).
\end{equation}
The shortcoming of using this approximation for calculating $\rho_t$
and $P_t$ are discussed well in literature 
\cite{Xu2009,Moustakidis12,Seif14}. In the following discussions,
the $a_{\rm sym}$, $L$ and $K_{\rm sym}$ denote their values at the
saturation density $\rho_0$, whereas, $a_{\rm sym,X}$, $L_{X}$ and
$K_{\rm sym,X}$ are  evaluated at $\rho= 0.X$ fm$^{-3}$.

To determine the fractions of the particles of  different species in
the NS matter the following constraints are used:
  \begin{itemize}
\item Balance equations for chemical potentials in the NS matter,
\bea
\mu_{n}&=& \mu_{p}+\mu_{e},\nonumber\\
\mu_{e} &=& \mu_{\mu}.
\label{EqCP}
\eea
\item Charge neutrality
\be
\rho_e +\rho_{\mu}=\rho_p.
\ee
\item Conservation of total baryon density 
\be
\rho =\rho_n+\rho_p.
\label{Eq:rhoB}
\ee
\end{itemize}
\begin{figure}
\centerline{\psfig{file=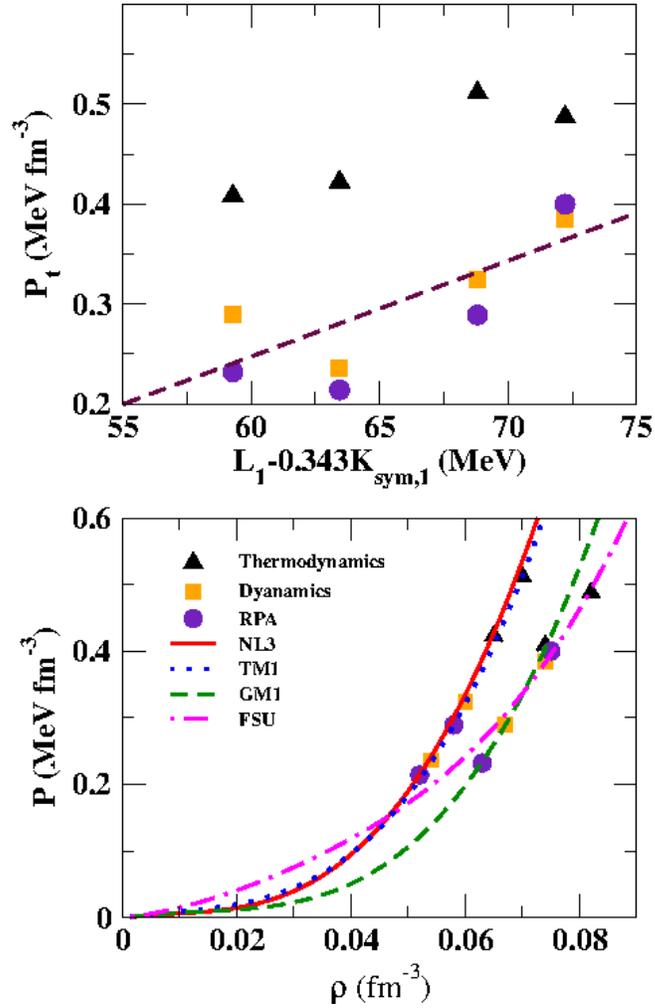,width=8.5cm}}

\caption{Plots for the EOSs for the $\beta$-equilibrated  matter
obtained
using  the NL3, FSU, TM1 and GM1 parameterizations of the RMF model
(lower panel).  The filled circles represent core-crust transition
density $\rho_t$ and the corresponding pressure  $P_t$ calculated using
RPA method. For the sake of comparison, the values of $\rho_t$ and
$P_t$  obtained using thermo-dynamical (filled triangles) and dynamical
methods(filled squares) of Ref.~\cite{Ducoin2011} are also shown. The
$P_t$ as a function of $L_{1}-0.343 K_{\rm sym,1}$ are  plotted in
the upper panel, where, $L1$ and $K_{\rm sym,1}$ represent the values
of values of $L$ and $K_{\rm sym}$ at $\rho = 0.01\text{ fm}^{-3}$,
respectively. The dash line is taken from Ref.~\cite{Ducoin2011}.}
\label{Fig1} \end{figure}

It is instructive to compare the core-crust
transition properties calculated within the RPA method with those obtained
from commonly used  thermo-dynamical and dynamical methods.  We plot
in Fig. \ref{Fig1}, the low density behavior for the EOS for
$\beta$-equilibrated  matter
(lower panel) obtained using NL3, FSU, TM1 and GM1 parameterizations
of the RMF model.  The solid symbols mark the values of the core-crust
transition  density $\rho_t$ and the corresponding pressure $P_t$ which
are obtained using dynamical (squares), thermo-dynamical (triangles)
and RPA (circles) methods.  The values of $\rho_t$ and $P_t$ calculated
within the RPA method seem to be  close to the ones obtained within the
dynamical method. The values of $\rho_t$ and $P_t$ calculated using the
thermo-dynamical method are somewhat higher.  The values of $P_t$ are
plotted as a function of $L_{1} - 0.343K_{\rm sym,1}$ in the upper panel.
The dash line is taken from Ref.~\cite{Ducoin2011}, which is obtained by
using the values of $P_t$ calculated from dynamical method.  It can be
seen that our values of $P_t$ calculated within the RPA method are more or
less consistent with  the linear correlation as shown by the dashed line.

\begin{figure}
\centerline{\psfig{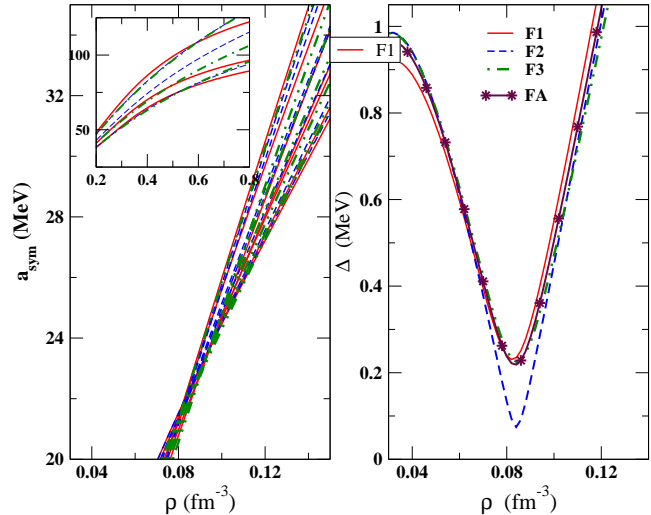}}
\caption{The density dependence of the symmetry energy coefficient $a_{\rm
sym}(\rho)$ for various ERMF models (left panel) and the variance $\Delta$
calculated using Eq. (\ref{eq:var}) (right panel). The labels F1, F2 and
F3 represent three different families of the ERMF models, whereas, FA
correspond to the results obtained by combining all the three families.
In the inset, the high density behaviour of $a_{\rm sym}(\rho)$ are
plotted for some selected forces from the F1,F2 and F3 families. }
\label{Fig2} \end{figure}

Before embarking on our main results,  let us look into the general
trends of the density dependence of symmetry energy for the various
ERMF models considered.  In the left panel of Fig. \ref{Fig2}, we plot
the symmetry energy as a function of density. 
For the sake of completeness, the high density behaviour
of $a_{\rm sym}(\rho)$ are plotted for some selected forces from the
F1,F2 and F3 families is plotted  in the inset.  These selected forces
correspond to the neutron-skin thickness around 0.16, 0.22 and
0.28 fm in $^{208}$Pb nucleus.
The variance $\Delta$
for $a_{\rm sym}(\rho)$ is plotted in the right panel.  The labels $F1$,
$F2$ and $F3$ denote three different families, while, $FA$ corresponds
to the results obtained by combing all the three families.
The values of $a_{\rm sym}$ at densities
in the range of $0.08 - 0.09 \text{ fm}^{-3}$ seems to be  more or less
same for all the different models.  The variance  $\Delta$  at a given
density is obtained using,
  \begin{equation}
\label{eq:var} \Delta^2=\frac{1}{n}\sum(a_{\rm sym}(\rho)-\bar{a}_{\rm
sym}(\rho))^2
  \end{equation} where, $n$ is the number of models and $\bar{a}_{\rm
sym}(\rho)$ is the average value at a density $\rho$.  The variance has
a minimum at $\rho \approx 0.08 \text{fm}^{-3}$ which is smaller
than $0.11 \text{fm}^{-3}$ as  obtained for a set of SHF and RMF  forces
\cite{Ducoin2011}.

\begin{figure}
\centerline{\psfig{file=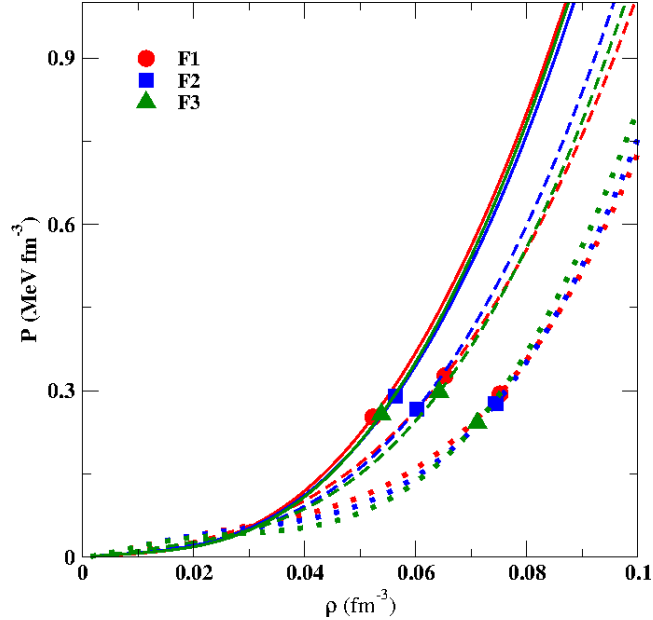,width=8.5cm}}
\caption{Plots for the EOSs for the $\beta$-equilibrated } matter
in terms of pressure verses density for the three different families of
the ERMF model: F1 (red), F2 (blue) and F3 (green).  The solid symbols
mark the values of transition density $\rho_t$ and the corresponding
pressure $P_t$ calculated within the RPA method.  For the clarity,
the results for only three forces for each of the families are plotted:
BSR1, BSR8, BSR15 (dotted), BSR4, BSR11, BSR18 (dashed) and BSR7, BSR14,
BSR21 (solid) lines (see text for details). \label{Fig3} 
\end{figure}

We now consider the core-crust transition properties obtained using three
different families of the ERMF models. In Fig. \ref{Fig3}, we display the
low density behavior of the EOS for $\beta$-equilibrated matter  for these ERMF models.
For the sake of clarity we plot the results only for a  few selected
forces for each of the families.   The dotted lines correspond to the
results for the BSR1, BSR8 and BSR15 forces belonging to the F1 (red),
F2 (blue) and F3 (green) families. Likewise, the dashed
correspond to  BSR4, BSR11, BSR18 and solid lines are for BSR7, BSR14,
BSR21.  All the dotted, dashed and solid lines correspond to the forces
associated with  the neutron-skin thickness in $^{208}$Pb to be around
0.16, 0.22 and 0.28 fm, respectively.  At low densities, $\rho \sim
0.03$ fm$^{-3}$, the behavior of the EOS is more or less independent
of the choice of the model and neutron-skin thickness in $^{208}$Pb.
With further increase in the density, the EOS show stronger dependence
on the choice of the neutron-skin thickness. For instance, BSR1, BSR8 and
BSR15 (dotted lines) correspond to the different families but have almost
the similar values of $a_{\rm sym}$, $L$ and  neutron-skin thickness.
We see that the EOSs associated with similar neutron-skin thickness
depend weakly on the  choice of the families of the models.  One may thus
expect the values of $\rho_t$ and $P_t$ depend not only on the $a_{\rm
sym}$ and $L$, but, also on the choice of the models.  In other words,
the core-crust transition properties may show some model dependence in
addition to their dependence on the symmetry energy parameters $a_{\rm
sym}$ and $L$.

\begin{figure}
\centerline{\psfig{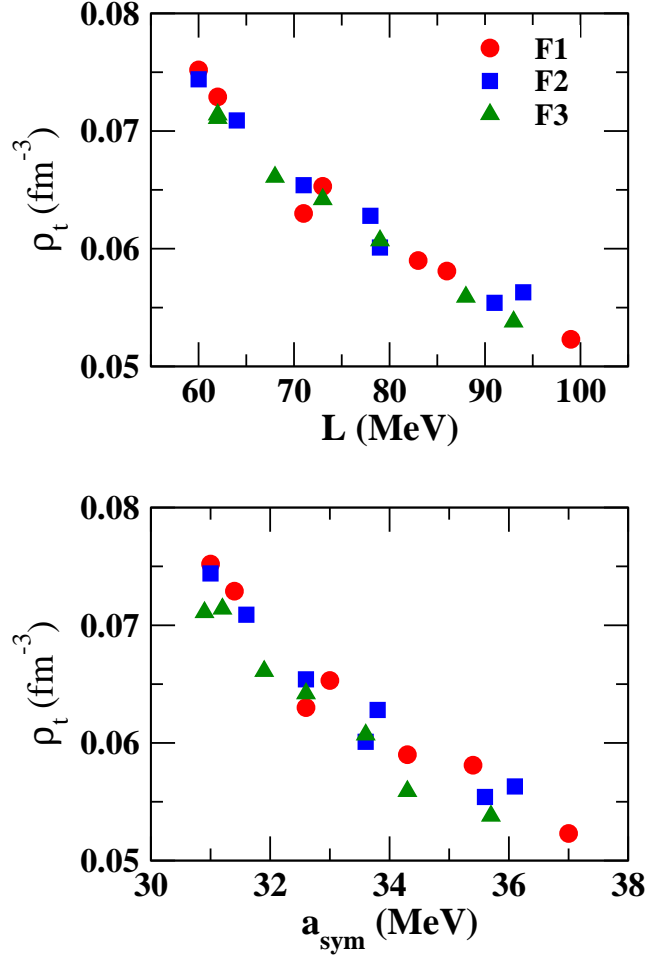}}
\caption{Plots for the $\rho_t$ for NS matter  as a function
of symmetry energy $a_{\rm sym}$ (lower panel) and its slope parameter $L$
(upper panel) for 3 different families of the ERMF models.}
\label{Fig4} \end{figure}

\begin{figure}
\centerline{\psfig{file=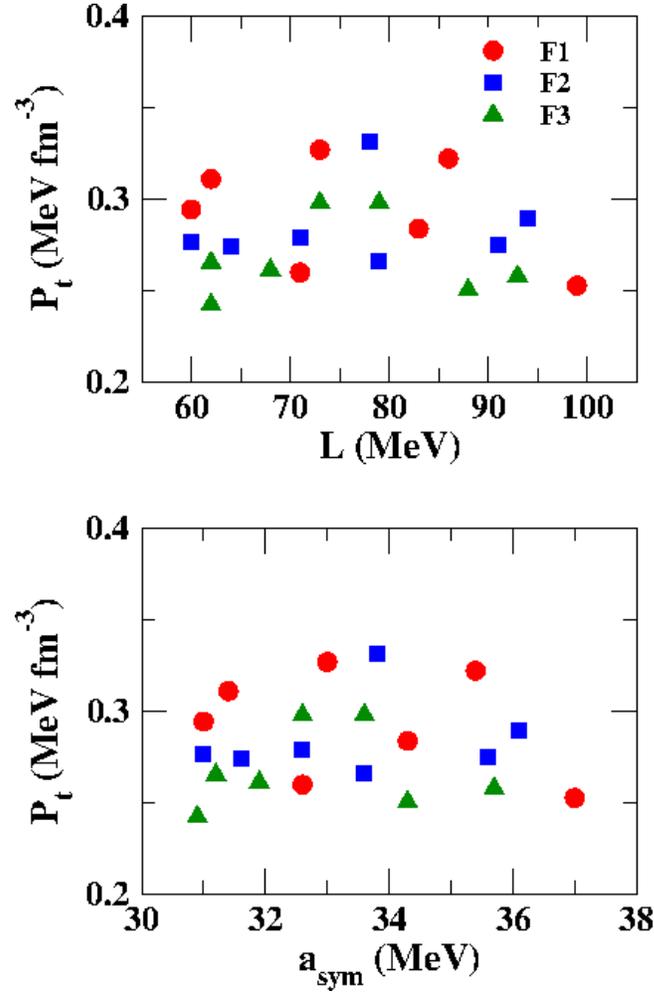,width=8.5cm}}
\caption{Plots for the $P_t$, for NS matter,  as
a function of $a_{\rm sym}$ (lower panel) and $L$ (upper panel) 
for 3 different families of
the ERMF models.  } \label{Fig5} 
\end{figure}

The values of $\rho_t$ obtained using ERMF models are
shown in Fig. \ref{Fig4} as a function of $a_{\rm sym}$ (lower panel) and
slope parameter $L$ (upper panel).  The values of $\rho_t$ are correlated
with the $a_{\rm sym}$ and $L$.  The $\rho_t - L$ correlations is stronger
than the $\rho_t - a_{\rm sym}$ correlations. The $\rho_t - a_{\rm sym}$
correlation is stronger within the same family.  But, $\rho_t - L$
correlations are almost model independent.  This is in conformity with
the earlier works \cite{Mousta2010,Ducoin2011}.  In Fig. \ref{Fig5}, we
plot $P_t$ for NS matter as a function of $a_{\rm sym}$ (lower) and $L$
(upper) panels for the ERMF models.  Our results indicate $P_t$ is not
well correlated with  $a_{\rm sym}$ and $L$. Consequently,  $P_t$ is
not correlated with $\rho_t$.  For the completeness, we list the values
of correlation coefficient for $\rho_t$ and $P_t$ with $a_{\rm sym}$,
$L$ and $K_{\rm sym}$ in Table \ref{tab:tab2}.  The $a_{\rm sym}$,
$L$ and $K_{\rm sym}$ refer to their values at the saturation density.
It is little too surprising that the $P_t - L$ correlations are weak
even within the same family of the ERMF model, though, the various
forces within the same family differ only in the density dependence of
the $a_{\rm sym}(\rho)$. On the other hand, the calculations in Refs.
\cite{Mousta2010,FP2012} based on a single model yield strong $P_t -
L$ correlations.

\begin{figure}
\centerline{\psfig{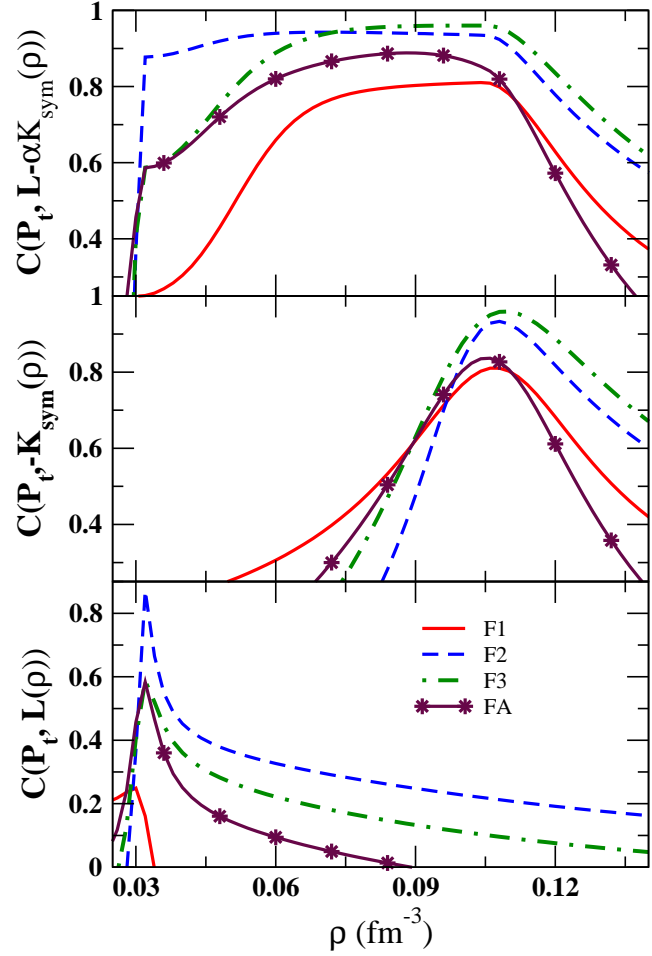}}
\caption{The correlation coefficients  for the pressure
$P_t$ with $L(\rho)$ (lower panel), $K_{\rm sym}(\rho)$ (middle panel)
and the linear combination $L(\rho) - \alpha K_{\rm sym} (\rho)$ (upper
panel) as a function of the density for 3 different families of
the ERMF models.} \label{Fig6} \end{figure}
We now explore the possibility of existence of strong correlations of
$P_t$  with $L(\rho)$, $K_{\rm sym}(\rho)$ and the linear combination
$L(\rho)- \alpha K_{sym}(\rho)$ at sub-saturation densities
($\rho<\rho_0$).  For the quantitative assesment, the Pearson's
correlation coefficients between $P_t$ and the  various symmetry energy
parameters  are calculated as a function of density. Pearson's correlation
coefficient $C(a,b)$ for a pair of variables $a$ and $b$ calculated for
$n$ number of  different models is given as,
\begin{equation}
C(a,b)=\frac{\sigma_{ab}}{\sqrt{\sigma_{aa}\sigma_{bb}}}
\end{equation}
where,
\begin{eqnarray}
\sigma_{ab} &=&\frac{1}{n}\sum_i a_i b_i -\left(\frac{1}{n}\sum_i a_i\right
)\left(\frac{1}{n}\sum_i b_i\right ).
\end{eqnarray}
The values of $C(a,b)$ lie in the range of $-1$ to $1$.  If $|C(a,b)|
$ = 1 then, the variables $a$ and $b$ are fully linearly correlated,
where as $C(a,b) = 0$ means, variables $a$ and $b$ are uncorrelated or
statistically independent.

\begin{figure}
\centerline{\psfig{file=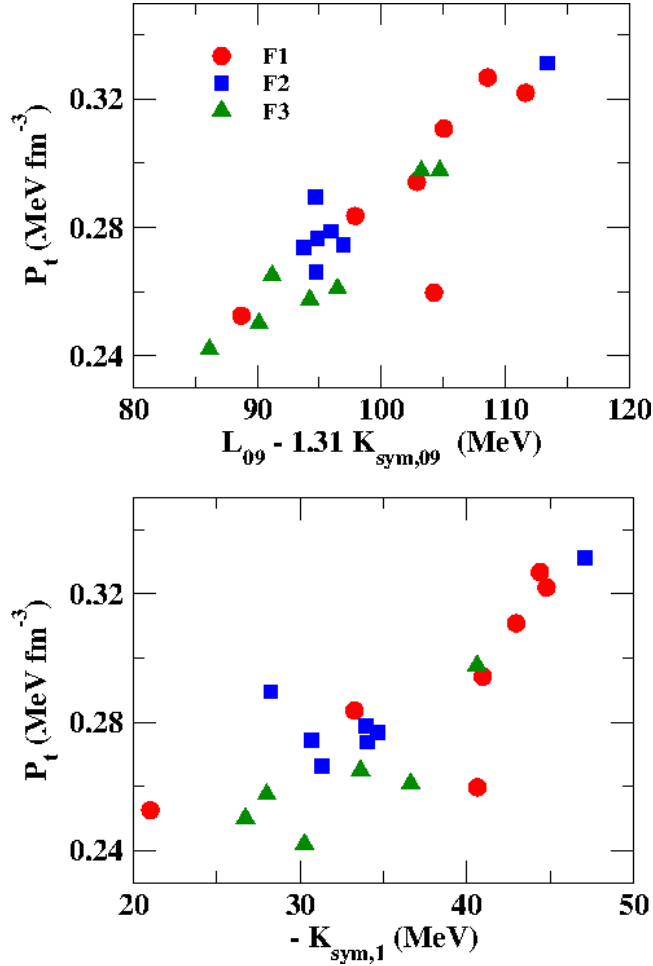,width=8.5cm}}
\caption{Plots for the pressure, $P_t$, at the transition
density as a function of $K_{\rm sym,1}$ (lower panel) and $L_{09}-1.31
K_{\rm sym,09}$ (upper panel) for 3 different families of the ERMF
models.} \label{Fig7} \end{figure}

It has been suggested in Ref. \cite{Ducoin2011}
that $P_t$ is reasonably correlated with the linear combination of
$L(\rho)$ and $K_{\rm sym}(\rho)$, with $\rho
=0.1\text{fm}^{-3}$.
In Fig. \ref{Fig6} the correlations of $P_t$ with $L(\rho)$
(lower), $K_{\rm sym}(\rho)$ (middle) and $L(\rho)- \alpha K_{sym}(\rho)$
(upper panels) are plotted as a function of density.  The value of
$\alpha$ is adjusted at a given density to maximize the correlation
coefficient.  The $P_t - L$ correlations are strong at quite low
densities for the F2 family of the ERMF models.  These correlations
become weak, once the results from all the different families of the models
are combined.  The $P_t$ seems reasonably correlated with $K_{\rm
sym}(\rho)$ as well as $L(\rho)-\alpha K_{\rm sym}(\rho)$ at some
sub-saturation densities.  The correlation coefficient $C(P_t, -K_{\rm
sym})$ peaks at $\rho = 0.1\text{ fm}^{-3}$, whereas, $C(P_t, L-\alpha
K_{\rm sym})$ peaks at $\rho = 0.09 \text{ fm}^{-3}$ for $\alpha = 1.31$.
It is to be noted that the peaks in plots for the $C(P_t, L -\alpha K_{\rm
sym})$ verses $\rho$ are much wider than those for the  $C(P_t, -K_{\rm
sym})$ verses $\rho$.  The strong correlations of $P_t$ with $K_{\rm
sym}$ at $\rho = 0.1$ fm$^{-3}$  may be due the use of systematically
varied models.  In Fig.  \ref{Fig7}, we plot the values of $P_t$ verses
$K_{\rm sym}$ (lower panel) and $P_t$ verses $L-1.31K_{\rm sym}$  (upper
panel), the values of $ L$ and $K_{sym}$ are evaluated at the densities
for which the $C(P_t, -K_{\rm sym})$ and $C(P_t, L-\alpha K_{\rm sym})$
correspond to their maximum values.  Our results for the correlations of
$P_t$ with the linear combination of $L$ and $K_{\rm sym}$ agree only
qualitatively with the ones obtained in Ref. \cite{Ducoin2011}. Our
values for the   correlation coefficient $C(P_t, L(\rho)-\alpha K_{\rm
sym}(\rho))$ is maximum at $\rho = 0.09 \text{fm}^{-3}$, while, its value
at $\rho = 0.1 \text{fm}^{-3}$ is significantly smaller than those of
Ref. \cite{Ducoin2011}. We also observe that the variance or the spread
in the values of  $a_{\rm sym}(\rho)$ for the ERMF models considered
in the present work is minimum at $\rho \sim 0.08 \text{fm}^{-3}$ (see
Fig. 2). Whereas, the variance of $a_{\rm sym}(\rho)$ for the set of SHF
and RMF forces employed in Ref. \cite{Ducoin2011} is minimum around $\rho
= 0.11 \text{fm}^{-3}$.  Thus, it seems that the $P_t$ is not correlated
with the symmetry energy parameters in a model independent manner.

\begin{table}[t]
\tbl{\label{tab:tab2}
The values of  correlation coefficients $C(A,B)$  with $A$ and $B$ being
the core-crust transition density $\rho_t$, corresponding pressure $P_t$
and various symmetry energy parameters at the saturation density.}
{\begin{tabular}{ccccc}
\hline
    &       F1  &      F2  &   F3 &FA  \\
\hline
$C(\rho_t,L)$    &-0.970   &-0.975   &-0.994  &-0.975\\
$C(\rho_t,a_{sym})$      &-0.963 & 0.966&-0.985& -0.954\\
$C(\rho_t,K_{sym})$      & 0.645 & 0.879  &0.813 & 0.643 \\
$C(P_t,L)$               &-0.363 & 0.157  &0.049& -0.065\\
$C(P_t,a_{sym})$         & -0.317 & 0.208  &0.090 & 0.017 \\
$C(P_t,K_{sym})$         & -0.355  & -0.545   & -0.624  & -0.130     \\
$C(P_t,\rho_t)$          & 0.416 & -0.108 & -0.071 & 0.128 \\

\hline
\end{tabular}}
\end{table}

\begin{figure}
\centerline{\psfig{file=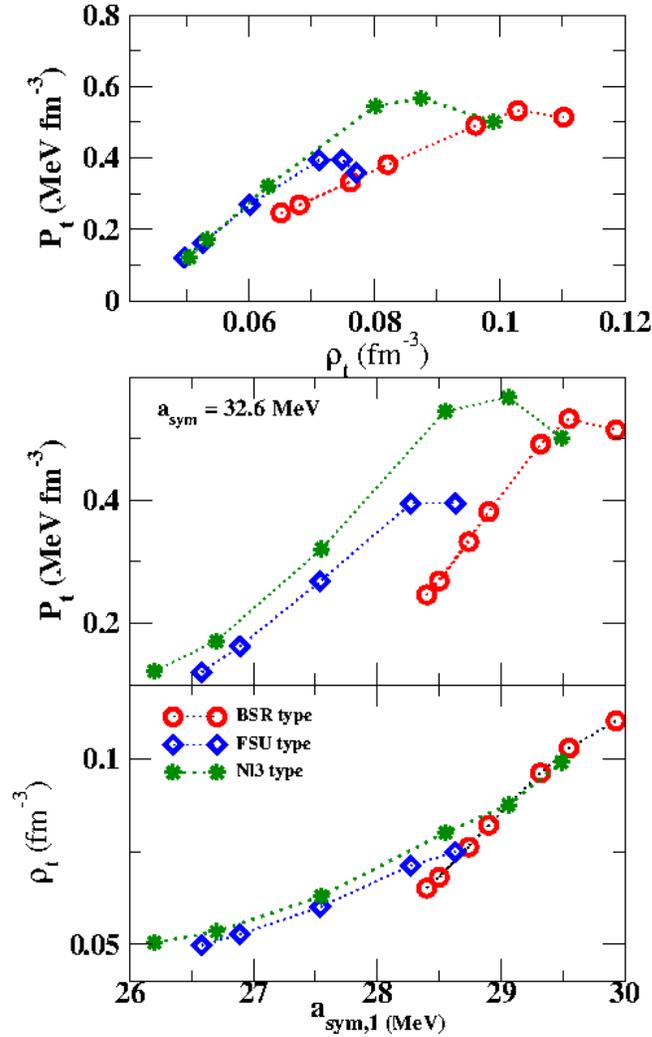,width=8.5cm}}
\caption{Plots for the $\rho_t$ (lower panel)
and $P_t$  (middle panel) as a function of $a_{\rm sym,1}$ and $P_t$
verses $\rho_t$ (upper panel) that obtained using BSR1, FSU and NL3 type
functional for the  RMF model. The value of $a_{\rm sym,1}$ is varied
at fixed  $a_{\rm sym} = 32.6$ MeV.}  \label{Fig8}
 \end{figure}

\begin{figure}
\centerline{\psfig{file=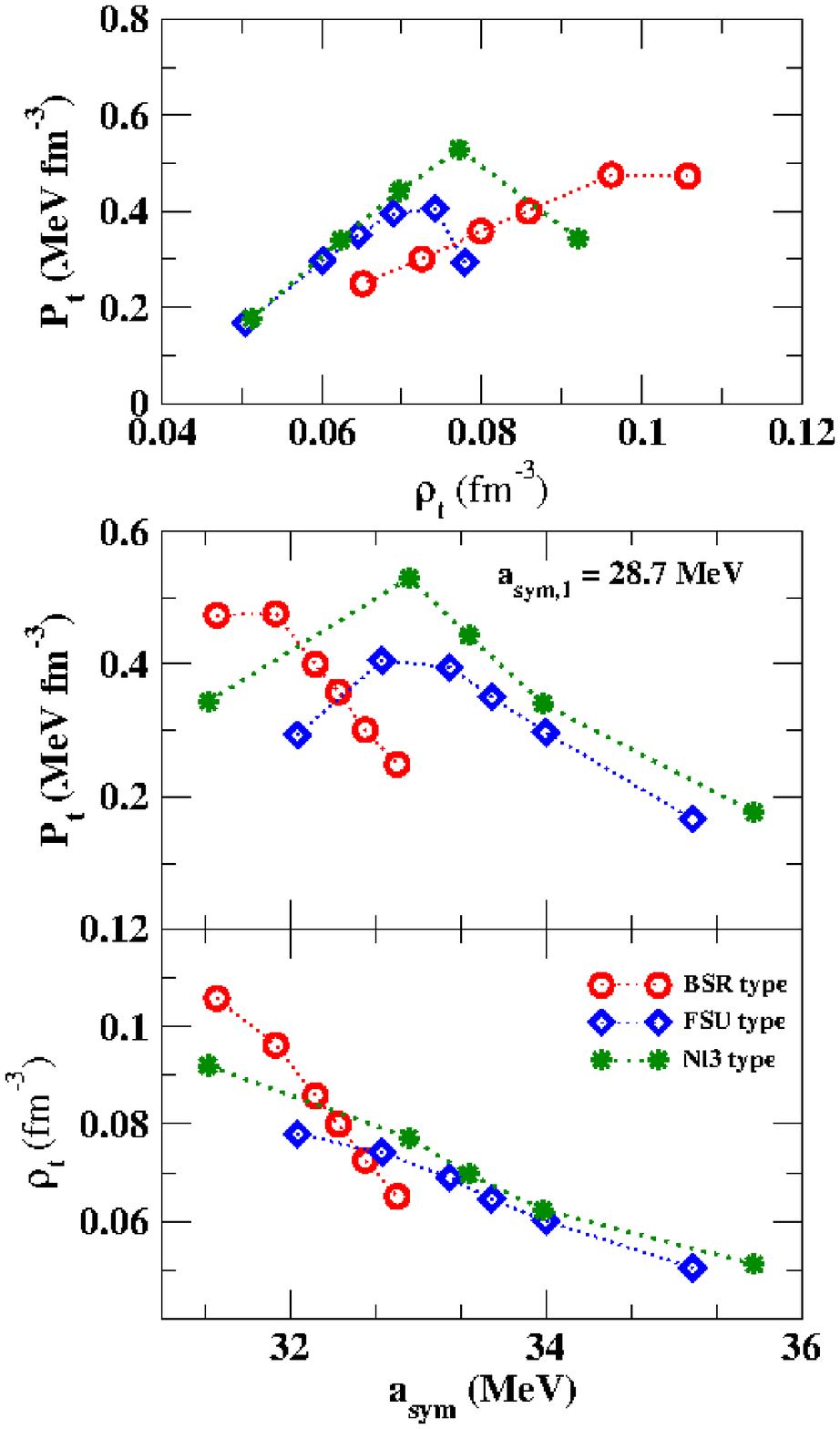,width=8.5cm}}
\caption{Same as that of \ref{Fig8}, but, $a_{\rm sym}$
is varied at fixed $a_{\rm sym,1} = 28.7$ MeV.} \label{Fig9}
 \end{figure}

So far we have studied the correlations of $\rho_t$ and $P_t$ with 
$a_{\rm sym}$, $L$ and $ K_{\rm sym}$
using different families of parameterizations  of the ERMF model. Each
of the parameterizations  were obtained by fitting exactly same set of
the experimental data for the bulk properties of finite nuclei except
for the neutron-skin thickness. We shall now study the variations
of $\rho_t$ and $P_t$ with $a_{\rm sym}$ and $a_{\rm sym,1}$ within
a single model  as it was done for  DD-PC1 model \cite{Mousta2010}.
For  Our investigation, we have considered   the  BSR1, NL3 and FSU type
of functionals for the RMF model.  The desired values of $a_{\rm sym}$
and $a_{\rm sym,1}$ are obtained by adjusting the coupling parameters
$g_\rho$ and $\eta_{2\rho}$ appearing in Eqs. (\ref{eq:eq3}) and
(\ref{eq:NLnuclagMX}).  In Fig. \ref{Fig8}, we display the variations
of $\rho_t$ and $P_t$ with $a_{\rm sym,1}$ for a fixed $a_{\rm
sym}=32.6$ MeV for the BSR1, NL3 and FSU parameterizations. Similarly,
in Fig. \ref{Fig9},  the variations of $\rho_t$ and $P_t$ with $a_{\rm
sym}$ for a fixed $a_{\rm sym,1} = 28.7$ MeV are displayed. 
In Table \ref{tab:tab3}, we list the values of the correlation coefficient
for $\rho_t$ and $P_t$ with $a_{\rm sym}$, and $a_{\rm sym,1}$.  It is
evident from the Figs. \ref{Fig8} and \ref{Fig9} and Table \ref{tab:tab3}
that the $\rho_t$ is correlated with $a_{\rm sym}$ as well as with
$a_{\rm sym,1}$ , irrespective of the model used.  Whereas $P_t$ is
strongly correlated only  with $a_{\rm sym,1}$ in a model  independent
manner. The $P_t - a_{\rm sym}$ correlations are model dependent.
For instance, the value of $\left|C(P_t, a_{\rm sym})\right| \sim 0.95$
for the BSR1 type of model which reduces to $\sim 0.6$ for the NL3 and
FSU type of models.  We can thus say once again that the pressure at the
transition density is correlated with  the symmetry energy parameter only
at  some sub-saturation density.  We have also repeated our calculations
for the variations of $\rho_t$ and $P_t$ with $a_{\rm sym} (a_{\rm sym,1})$
by fixing $a_{\rm sym,1} (a_{\rm sym})$ to different values. The results
are qualitatively the same; $P_t - a_{\rm sym}$ correlations are model
dependent.

\begin{table}[t]
\tbl{ Values for the various correlation coefficients obtained
by varying $a_{\rm sym}$ or $a_{\rm sym,1}$  within a single model.
Three different models, BSR1, FSU and NL3, are considered. The values
of $a_{\rm sym}$ is varied by fixing $a_{\rm sym,1}= 28.7 \text{ MeV}$ ,
while, $a_{\rm sym,1}$ is varied by fixing $a_{\rm sym}=32.6\text{ MeV}$.
The values of correlation coefficients obtained by combining the results
from all the three models are presented in the last column.}
{ \begin{tabular}{ccccc}
\hline
    &       BSR1  &      FSU  &   NL3 & All  \\
\hline
  & & $a_{sym} $= 32.6 MeV\\
\hline
$C(\rho_t,a_{sym,1})$      & 0.996  & 0.997  & 0.988  &  0.942 \\
$C(P_t,a_{sym,1})$         & 0.956  & 0.947  & 0.952  &  0.879  \\
$C(P_t,\rho_t)$          & 0.979  & 0.960  & 0.914  & 0.906    \\
\hline
  & & $a_{sym,1} $= 28.7 MeV \\
\hline
$C(\rho_t,a_{sym})$      & -0.995 & -0.995  &-0.983   & -0.917 \\
$C(P_t,a_{sym})$         & -0.946 & -0.659  &-0.570   & -0.612  \\
$C(P_t,\rho_t)$          &  0.973 & 0.704   & 0.543   &  0.703    \\
\hline
\label{tab:tab3}
\end{tabular}}
\end{table}

\section{Conclusions}
\label{conclu}
The variations of  core-crust transition properties  in the neutron star
with symmetry energy parameters are investigated using three different
families of the systematically varied  ERMF model.  These families
of the ERMF model mainly differ in the choice of the strength for the
$\omega$-meson self-coupling.  Several parameterizations for each of
the families are so considered that they yield  wide variations in the
density dependence of the symmetry energy.

Our results indicate that the transition density $\rho_t$ is
strongly correlated with the symmetry energy slope parameter $L$ at
the saturation density which is in harmony with the earlier studies
\cite{Mousta2010,Ducoin2011}.  The $\rho_t$ is also correlated with
the symmetry energy at the saturation density, but, the correlations
are marginally model dependent.  The pressure $P_t$ at the transition
density, however,  does not show any meaningful correlations with the
values of various symmetry energy parameters at the saturation density.
The possibility of existence of strong correlations between the pressure
at the transition point and the symmetry energy parameters evaluated
at the sub-saturation density are explored.  It is found that  $P_t$
is better correlated with the curvature parameter $K_{\rm sym}$ alone
or with the linear combination of $L$ and $K_{\rm sym}$, both the
quantities calculated at some sub-saturation density. We observe
that the density $\rho= 0.09 \text{ fm}^{-3}$ at which the correlation
coefficient $C(P_t, L(\rho)-\alpha K_{\rm sym}(\rho))$ peaks is quite
close to the one  at which the variance of $a_{\rm sym}(\rho)$ is minimum.
The strong correlations between $P_t$ and linear combination of $L(\rho)$
and $K_{\rm sym}(\rho)$   at $\rho = 0.1\text{ fm}^{-3}$ for a set of
SHF and RMF models \cite{Ducoin2011} for which the  variance of $a_{\rm
sym}(\rho)$ is minimum at $\rho = 0.11 \text{fm}^{-3}$ also supports our
observation.  Though, the pressure at the transition point is correlated
with linear combination of the symmetry energy slope and the curvature
parameters evaluated at a sub-saturation density, such correlations show
some degree of model dependence.

We also study the dependence of core-crust transition properties
on various symmetry energy parameters using a single model. In this case,
the symmetry energy parameters are varied by modifying the values of
the model parameters around their optimal values.
Two different kinds of variations in the  symmetry energy parameter
$a_{\rm sym}$ are considered.  The values of $a_{\rm sym}$ are varied
at the saturation density by keeping its value fixed at the density
$\rho = 0.1\text{ fm}^{-3}$. Another type of variations in $a_{\rm
sym}$ is obtained by changing its value at the density $\rho = 0.1
\text{ fm}^{-3}$, but, keeping it fixed at the saturation density.
The calculations are performed for the BSR1, NL3 and FSU type of the
RMF model.  The transition density is found to be  strongly correlated
with the values of $a_{\rm sym}$ calculated at the saturation density
as well as those at $\rho = 0.1 \text{ fm}^{-3}$, irrespective of the
model used. The pressure at the transition density is correlated in the
model independent manner only with the $a_{\rm sym}$ at the $\rho =
0.1 \text{fm}^{-3}$.  The correlations of pressure at the transition
density with the $a_{\rm sym}$ at the saturation density are highly
model dependent.  It thus appears once again that the pressure at the
transition density is  at best correlated with symmetry energy parameters
at some sub-saturation density.

\section*{Acknowledgements}
Anto Sulaksono acknowledges the support given by Universitas Indonesia.


\begin{thebibliography}{160} 
\bibitem{Lattimer2012}J. M. Lattimer and Y. Lim, 
\Journal{\APJ}{771}{51}{2013}.
\bibitem{Fur2002} R. Furnstahl,
\Journal{\NPA}{706}{85}{2002}.
\bibitem{Steiner2005}A. W. Steiner, M. Prakash, J. Lattimer, and P. J. Ellis,
\Journal{\PRpt}{411}{325}{2005}.
\bibitem{Warda2009}M. Warda, X. Vinas, X. Roca-Maza, and M.Centelles,
\Journal{\PRC}{80}{024316}{2009}.

\bibitem{Cente2009} M.Centelles, X. Roca-Maza, X. Vinas, and M. Warda,
\Journal{\PRL}{102}{122502}{2009}.
\bibitem{Pie1} J. Piekarewicz, B. K. Agrawal, G. Colo, W. Nazarewicz,
N. Paar, P.-G. Reinhard, X. Roca-Maza, and D. Vretenar,
\Journal{\PRC}{85}{041302}{2012}.

\bibitem{Agra2012}B. K. Agrawal, J. N. De, and S. K. Samaddar,
\Journal{\PRL}{109}{262501}{2012}.
\bibitem{Roca-Maza13} X. Roca-Maza, M. Brenna, B. K. Agrawal, P. F.
Bortignon, G. Colo, Li-Gang Cao, N. Paar, and D. Vretenar,
\Journal{\PRC}{87}{034301}{2013}.

\bibitem{Mousta2010}Ch. C. Moustakidis, T. Nik\v{s}i\'c, G. A. Lalazissis, D. Vretenar, and P. Ring,
\Journal{\PRC}{81}{065803}{2010}.
\bibitem{FP2012}F. J. Fattoyev and J. Piekarewicz, 
 \Journal{\PRC}{86}{015802}{2012}.
\bibitem{Xu2009}J. Xu, L-W. Chen, B-A. Li and H-R. Ma,
\Journal{\APJ}{697}{1549}{2009}.
\bibitem{Ducoin2011}C. Ducoin, J. Margueron, C. Provid\^encia, and I. Vida\~na,
\Journal{\PRC}{83}{045810}{2011}.
\bibitem{Newton2013}W. G. Newton, M. Gearheart and Bao-An Li,
\Journal{\APJS} {204}{9}{2013}.
\bibitem{Klupfel09}P. Kl\"upfel and P.-G. Reinhard and T. J. Burvenich and  J. A. Maruhn 
\Journal {\PRC}{79}{034310}{2009}.
\bibitem{BA2010} B. K. Agrawal,
\Journal{\PRC}{81}{034323}{2010}.
\bibitem{DKA2007} S. K. Dhiman, R. Kumar, and B. K. Agrawal,
 \Journal{\PRC}{76}{045801}{2007}.
\bibitem{KS2009} A. Sulaksono and Kasmudin,
\Journal{\PRC}{80}{054317}{2009}; Kasmudin and A. Sulaksono,
\Journal{\IJMPE}{20}{1271}{2011}.
\bibitem{Kubis2007} S. Kubis,
\Journal{\PRC}{76}{025801}{2007}; \Journal{\PRC}{70}{065804}{2004}.
\bibitem{Lattimer2007}J. M. Lattimer and M. Prakash,
\Journal{\PRpt}{333}{121}{2007};\Journal{\APJ}{550}{426}{2001}.
\bibitem{Li2002}Bao-Ann Li, A. T. Sustich, M. Tilley and B. Zhang,
\Journal{\NPA}{699}{493}{2002}.
\bibitem{Nilsen1991}M. Nielsen, C. Provid\^encia and J. da Provid\^encia,
\Journal{\PRC}{44}{209}{1991}.
\bibitem{Provindencia2006} C. Provid\^encia, L. Brito, S. S. Avancini,
D. P. Menezes and Ph. Chomaz, \Journal{\PRC}{73}{025805}{2006}.
\bibitem{Avancini2010} S. S. Avancini, S. Chiacchiera, D. P. Menezes and C. Provid\^encia,
\Journal{\PRC}{82}{055807}{2010}.
\bibitem{Pathick1995}C. J. Pethick, D. G. Ravenhall and C. P. Lorenz,
\Journal{\NPA}{584}{675}{1995}.
\bibitem{Douchin2000}F. Douchin and P. Haensel,
\Journal{\PLB}{485}{107}{2000}.
\bibitem{Ducoin2007}C. Ducoin,  Ph. Chomaz and F. Gulminelli,
\Journal{\NPA}{789}{403}{2007}.
\bibitem{Lim1989}K. Lim and C. J. Horowitz,
\Journal{\NPA}{501}{729}{1989}.
\bibitem{HP2001} C. J. Horowitz and J. Piekarewicz,
\Journal{\PRL}{86}{5647}{2001}.
\bibitem{Carr} J. Carriere, C. J. Horowitz, and J. Piekarewicz,
\Journal{\APJ}{593}{463}{2003}.
\bibitem{AS2009}A. Sulaksono, T. J. B\"urvenich, P.-G. Reinhard and J. A. Maruhn,
\Journal{\PRC}{79}{044306}{2009}; A. Sulaksono, T. Mart, T. J. B\"urvenich and J. A. Maruhn,
\Journal{\PRC}{76}{041301(R)}{2007}.
\bibitem{AM2006}A. Sulaksono and T. Mart,
\Journal{\PRC}{74}{045806}{2006}.
\bibitem{MA2008}T. Mart and A. Sulaksono,
\Journal{\PRC}{78}{025808}{2008};
\bibitem{FST96}R. J. Furnstahl, B. D. Serot and H. B. Tang,
\Journal{\NPA}{598}{539}{1996}; \Journal{\NPA}{615}{441}{1997}.
\bibitem{BB77} J. Boguta and A. R. Bodmer,
\Journal{\NPA}{292}{413}{1977}.
\bibitem{Lala97}G. A. Lalazissis, J. K\"onig and P. Ring,
\Journal{\PRC}{55}{540}{1997}.
\bibitem{Bender} M. Bender, K. Rutz, P.-G. Reinhard, J. A. Maruhn and W. Greiner,
\Journal{\PRC}{60}{034304}{1999}.
\bibitem{Garg} U. Garg {\it et al.},
\Journal{\NPA}{788}{36}{2007}. 
\bibitem{Li07}T. Li {\it et al.},
\Journal{\PRL}{99}{162503}{2007}.
\bibitem{Estal} M. Del Estal,  M. Centelles, X. Vi\~nas and S. K. Patra,
\Journal{\PRC}{63}{024314}{2001}.
\bibitem{Toki}Y. Sugahara and H. Toki,
\Journal{\NPA}{579}{557}{1994}.
\bibitem{Estal2} M. Del Estal,  M. Centelles and X. Vi\~nas,
\Journal{\NPA}{650}{443}{1999}.
\bibitem{Mulser96}H. Mueller and B. D. Serot, 
\Journal{\NPA}{606}{508}{1996}.
\bibitem{Fatt2012} F. J. Fattoyev and J. Piekarewicz,
\Journal{\PRC}{86}{015802}{2012}.
\bibitem{TSil2005} T. Sil, M. Centelles, X. Vi\~nas and J. Piekarewicz,
\Journal{\PRC}{71}{045502}{2005}.
\bibitem{Pieka2} B. G. Todd-Rutel and J. Piekarewicz,
\Journal{\PRL}{95}{122501}{2005}.
\bibitem{GM91} N. K. Glendening and S. A. Moszkowski,
\Journal{\PRL}{67}{2414}{1991}.
\bibitem{Typel99}  S.Typel and H. H. Wolter
\Journal{\NPA}{656}{331}{1999}.
\bibitem{Vretenar05} D. Vretenar, G. A. Lalazissis, T. Nik\v{s}i\'c, and P. Ring,
\Journal{Eur. Phys. J. A}{25}{555}{2005}.
\bibitem{Burvenich02}T. Burvenich, D. G. Madland, J. A. Maruhn, and
P.-G.Rei nhard
\Journal{\PRC}{65}{044308}{2002}.
\bibitem{Moustakidis12} Ch. C. Moustakidis
\Journal{\PRC}{86}{015801}{2012}.
\bibitem{Seif14} W. M. Seif and D. N. Basu
\Journal{\PRC}{89}{028801}{2014}.
\end{thebibliography}
\end{document}